# Multifunctional Meta-Optic Systems: Inversely Designed with Artificial Intelligence

*Dayu Zhu[1], Zhaocheng Liu[1], Lakshmi Raju[1], Andrew S. Kim[1], Wenshan Cai[1,2*]*

[1] School of Electrical and Computer Engineering, Georgia Institute of Technology, Atlanta, Georgia 30332-0250

[2] School of Materials Science and Engineering, Georgia Institute of Technology, Atlanta, Georgia 30332-0295

* Correspondence should be addressed to W.C. (wcai@gatech.edu)

**Abstract**

Flat optics foresees a new era of ultra-compact optical devices, where metasurfaces serve as the foundation. Conventional designs of metasurfaces start with a certain structure as the prototype, followed by an extensive parametric sweep to accommodate the requirements of phase and amplitude of the emerging light. Regardless of how computation-consuming the process is, a predefined structure can hardly realize the independent control over the polarization, frequency, and spatial channels, which hinders the potential of metasurfaces to be multifunctional. Besides, achieving complicated and multiple functions calls for designing a meta-optic system with multiple cascading layers of metasurfaces, which introduces super exponential complexity. In this work we present an artificial intelligence framework for designing multilayer meta-optic systems with multifunctional capabilities. We demonstrate examples of a polarization-multiplexed dual-functional beam generator, a second order differentiator for all-optical computation, and a space-polarization-wavelength multiplexed hologram. These examples are barely achievable by single-layer metasurfaces and unattainable by traditional design processes.

The metasurface is one of the most appealing ideas in nanophotonics in the last decade[1,2]. As the two-dimensional counterpart of metamaterials, metasurfaces introduce abrupt phase change to the interface, which empowers the ultrathin layer to function as a bulky optical component. This intriguing advantage allows for numerous applications of metasurfaces, such as metalenses, wavefront shaping, compact imaging, to name a few[3-6]. At the early stage, most optical metasurfaces were implemented through plasmonics[7], which can achieve the complete phase control of the cross-polarized output wave. Nevertheless, the theoretical limit of the cross-polarized conversion efficiency of plasmonic metasurfaces is only 25%, and the ohmic loss associated with plasmonic metasurfaces further leads to lower efficiency. The other family of metasurfaces, the dielectric ones, can achieve much higher efficiency and are better compatible with the modern semiconductor industry[8]. The transmittance of non-resonant dielectric metasurfaces can reach near unity in the optical regime, providing high efficiency but mediocre amplitude control. Apart from the individual merit and demerit of plasmonic and dielectric metasurfaces, most metasurfaces are built upon single-layer unit cells, which have rather low degrees of freedom (DOF) to accomplish a sophisticated objective. For instance, a single-layer metasurface can hardly be highly multifunctional, since the optical responses of a unit cell at different polarizations, frequencies, and spatial positions are correlated with each other. Thus, the polarization, frequency, and position cannot be used as independent channels to multiplex information. Trials are made to design multifunctional metasurfaces by integrating several types of nanostructures onto one metasurface[9,10], where each type of nanostructure functions towards a distinct objection. Although effective sometimes, this scheme may impede the resolution of metasurfaces and the crosstalk between nanostructures is unavoidable. These limitations of metasurfaces call for a meta-optic system with multiple metasurfaces to fulfill a complicated objective, analogous to an optical system with cascading lenses. In this sense, a meta-optic system is a favorable choice for implementing highly multifunctional devices.

The meta-optic system can be intuitively viewed as multiple layers of metasurfaces, with a subwavelength total thickness[11,12]. Nevertheless, most multilayer metasurfaces regard each layer as an individual device, and the ultimate performance can be calculated through ray-tracing or other methods. In contrast to the general idea of multilayer metasurfaces, here we introduce the multilayer meta-optic system, which does not necessarily require each layer to have a distinct function. Hence, we only need to design the entire system towards the ultimate objective without distinguishing the layer-wise properties. Besides, the meta-optic system also bridges the gap between plasmonic and dielectric metasurfaces. For one thing, the meta-optic system provides sufficient propagation length for waves, similar to dielectric metasurfaces, which aids the full coverage of phase. For another, the meta-optic system can be composed of either metals or dielectrics, or both, to achieve complete amplitude modulation. In addition, high DOF of the meta-optic system provides independent control of the optical responses of different multiplexed channels. The microscopic patterns of such meta-optic systems can be as complicated as arbitrary geometries, which further improves the DOFs. In this article, we will demonstrate that meta-optic systems can be one of the best candidates to achieve multifunctional purposes for light waves. As for novel applications such as all-optical computation, signal and imaging processing, pattern recognition, etc., meta-optic systems also represent one of the most promising media.

Meanwhile, increased design complexity is coexistent with the advantages of the meta-optic system. The conventional design process requires a predefined nanostructure as the building block, then a thorough sweep of the geometric parameters of the nanostructure is necessary. This design flow suffers two major shortcomings when applied to meta-optic systems. One is that the selected nanostructure may not cover the solution space of the desired functions. For instance, predefining a nanostructure with distinct functionality at different wavelengths is never a closed form question, and tedious trial-and-error attempts to vary the parameters of the nanostructure is usually necessary, but with no guarantee of a solution. The other is that adding layers will increase the number of parameters super exponentially. Therefore, the parameter-scanning process will consume huge computational resources and storage, which is unacceptable or even impractical. This dilemma can be perfectly mitigated by our artificial intelligence (AI) framework. Advances in deep learning and computational hardware have given birth to diverse applications of AI in information technology[13-15], such as pattern recognition, natural language processing, autonomous vehicles, etc. Besides, AI also benefits the methodology of natural science[16,17], including biology, chemistry, quantum mechanics, as well as optics and nanophononics. Recently, the interdisciplinary regime between optics and AI has caught intense attention. Various studies have validated the potential of leveraging AI for the design of optical components and proven that AI-design methods can essentially outperform traditional human-design processes on certain tasks[18-24]. It is worth noting that researchers up to date may have not paid enough attention to the tasks with formidably high complexity, which are hardly resolvable by conventional design processes and thus exclusive to AI-design methods. In this work, we introduce the first AI framework for the discovery and design of a multilayer multifunctional meta-optic system, which is too complicated to be accomplished through conventional design processes. As solid illustrations, we report three examples designed by our framework: a polarization-multiplexed dual-functional beam generator, a second order differentiator for all-optical computation, and a space-polarization-wavelength multiplexed hologram. These functions are barely achievable by single-layer metasurfaces, and the devices are hardly approachable by conventional design means. The designed devices are constituted of highly complicated patterns, which indicates the extremely high DOFs of the structures involved.

The design flowchart of our AI framework is presented in Figure 1. The algorithm architecture consists of a neural network (NN) generator, a NN simulator, and an Evolutionary Strategy (ES) based optimizer[25]. As for the mechanism of the AI framework, in simple terms, the NN generator decodes a batch of randomly initialized latent vectors and produces corresponding candidate nanostructures, followed by the NN simulator computing the transmittance and reflectance of those nanostructures. Next, the ES optimizer selects the structures with good performance, refines and evolves their latent vectors, and sends the latent vectors back to the generator to get a better batch of structures. The procedure proceeds as a loop until satisfactory structures are obtained. To be specific, the logic and the functionality of the NN generator, the NN simulator, and the ES optimizer are demonstrated sequentially.

The NN generator is the generative model of a generative adversarial network (GAN)[26]. Most GANs are constructed by the convolutional neural network (CNN)[15], which is beneficial to extract the inherent features of images. However, different nanostructures of metasurfaces seldom share common features, so we turn to a compositional-pattern generating network (CPPN)[27] to implement the generator of the GAN, which is proven to be effective for generating artistic patterns with various styles. As for the NN simulator, it is adapted from ResNet18 [15], with 64 by 64 pixelated images as the input and the complex transmission and reflection spectra (170 THz to 600 THz in frequency, corresponding to the wavelength from 500 nm to 1765 nm in free pace) as the output. Each image represents the two-dimensional nanostructure of a unit cell, with a unit cell size of 320 nm and pattern thickness of 40 nm. To be reasonable for multilayer designs, we assume all nanostructures are gold and embedded in a polymer with a refractive index of 1.57. Although the simulator is a very deep NN, it is trained with a tiny dataset of only 2,000 image-spectra pairs. We avoid simulating the spectra of a gigantic number of nanostructures to generate the training dataset by applying transfer learning to reduce the data requirement. In our previous research[25], we have simulated 16,000 image-spectra data pairs of unit cells of the same size and pattern thickness, where the gold structure is deposited on glass and exposed in air. Although the material configurations are different, transfer learning is capable of identifying the connection. We first use the previous data (air-glass-gold configuration) to train our simulator and obtain 98% accuracy. After that, with freezing all the network layers of the simulator except for the last layer, we continue training the simulator with only 2,000 new data (polymer-gold configuration) and achieve 96% accuracy. The high efficacy of the simulator suggests that transferring learning successfully transfers the knowledge from the air-glass environment to the polymer-embedded environment. It also indicates that our framework can be readily transferred into any other setups of materials or sizes.

With the generator and the simulator working on the unit cells of single layer nanostructures, the optimizer assembles unit cells into multilayer. The unit cells at the same planar position but on different layers can be collectively viewed as a supercell of the meta-optic system, and the overall performance of the supercell is calculated by the matrix-chain multiplication of the wave matrix of each unit cell and the spacers in between[11]. In one iteration, the ES optimizer will optimize one layer of the supercell and keep other layers unaltered. After the selection, reproduction, mutation and elimination processes of the evolutionary algorithm, the optimizer will select the nanostructure which leads to the least discrepancy between the optical response of the supercell and the predefined objective. In the next iteration, the framework will optimize the next layer of the supercell (from the top layer to the bottom, then back to top). The loop will proceed until satisfactory performance of the supercell is obtained, then the framework will work on designing the supercells at other planar positions. Compared to other optimization methods for multilayer metasurfaces, our ES optimizer needs no sub-objective of every layer but instead optimizes the supercell towards an overall objective, which is more efficient and avoids the null-space problem of the solutions to the sub-objectives. Further details of our AI algorithm framework are presented in the Supplementary Information.

The first case study of the AI-designed meta-optic system is a dual-functional beam generator, as shown in Figure 2. When the incident light is an *x*-polarized plane wave, the transmitted *x*-polarized output represents the zeroth order Bessel beam, while the *y*-in-*y*-out response is an Airy beam. Beam shaping is a classical application of metasurfaces, by which the ultrathin beam generator is feasible, while bulky optical components, such as axicons and lenses, are no longer necessary. Beam shaping calls for both the magnitude and the phase of the optical responses of the metasurface to be spatially distributed, which increases the complexity in designing the unit cells. Besides, to achieve distinct functions for different states of polarizations, the optical responses for both polarizations should be addressed. The task is barely feasible through traditional design processes, and tailoring a single-layered metasurface is not a viable route to achieving polarization-multiplexed amplitudes and phases. One promising solution is to implement the AI algorithm to design a multilayer meta-optic system with the target multifunctional properties. Without loss of generality, we preset the meta-optic system to possess three layers, with 25 spatially variant supercells in both *x* and *y* directions, and the polymer spacer between adjacent layers is set to 200 nm. The wavelength here is 659 nm in the polymer, which is randomly chosen to validate the performance of our algorithm. As for the design process, we first need to pin down the design objective. The expected Bessel and Airy beams are each with four lobes along each direction, providing sufficient resolution for the generated beam along the length of 25 unit cells. Upon settling down the spatial amplitude and phase distributions of Bessel and Airy beams, the framework will generate the best possible supercells at each planar position and the full-wave simulation will further be used to validate the design efficacy.

The details of the dual-functional Bessel-Airy beam generator are displayed in Figure 2. Figures 2a and 2c depict the detailed structure of each layer, from the top layer to the bottom, respectively, and we assume the incidence light illuminates from the top. Although some of the generated patterns are very complicated for actual device fabrication, we can apply denoising and Fourier filter algorithm to extract the main features of the patterns, which are much easier to fabricate. This practice will not comprise the device performance, as the filtered tiny details of the structures do not contribute much to the farfield distributions. The detailed discussion concerning the fabrication tolerance is provided in Supplementary Information. Figures 2d and 2g demonstrate the comparison between the amplitude distributions of designed and target transmittances for the two polarizations on the central axis of the beam generator, which reveals the remarkable agreement between the achieved and target performance. For the whole meta-optic system, the average design accuracy of the optical responses of the supercells can reach approximately 90%, which guarantees the effectiveness of the designed dual-functional beam generator. Figure 2e and 2h present the simulated output of the beam generator at *x* and *y* polarizations, respectively, where the light field distributions undoubtedly represent the desired Bessel and Airy patterns. The simulated light propagations are illustrated in Fig. 2d and 2e, where the non-diffracting characteristics of both beams, and the self-bending of the Airy beam are evidenced. Besides being multifunctional, the designed beam generator exhibits decent efficiency compared to beam generators by conventional plasmonic metasurfaces. For *x*-polarized incidence, assuming an input intensity of 100%, the peak intensity of the transmitted Bessel beam reaches 98%. For *y*-polarized incidence, since the peak value of an Airy function is not unity, to reveal the capability of our AI algorithm, we randomly choose 50% as the peak value of the objective Airy beam on the central axis, as indicated in Figure 2d. Our design fulfills the objective as expected, and the peak intensity of the transmitted Airy beam is approximately 55%. Compared to conventional plasmonic metasurfaces with a theoretical upper limit of 25% transmitted power, the designed meta-optic system achieves far superior efficiency.

The second example of the AI designed multilayer meta-optic system is a second-order differentiator, which is a representative example of all-optical computation and all-optical signal processing. Calculation and signal-processing with metasurfaces has long been an intriguing topic to achieve all-optical computation[12,28,29]. A limited number of approaches have been proposed on this subject, including the use of a meta-optic system with one metasurface sandwiched between two graded-index (GRIN) slabs, and the application of multilayer homogeneous metamaterials[28]. However, the first approach mentioned above requires design of a metasurface and two bulky GRIN structures independently. Meanwhile the second approach is only applicable to math operations with even symmetry and the design may have too many layers or call for unrealistic materials. Inspired by the theoretical background of the multilayer metamaterial method, we can exploit the meta-optic system to realize certain functions for all-optical computation. The meta-optic system is an ideal candidate to achieve the spatially variant Green's function, which represents wavenumber dependent transmittance in k-space. Since the spatially distributed supercells are not necessarily symmetric, the meta-optic system can be adapted to any linear operation, such as derivation, integration, convolution, etc.

In our example of meta-system for second-order differentiation, the schematic of the three-layered computational meta-optic system is illustrated in Fig. 3a. In this showcase, when the incident light (659 nm in the polymer) is *y*-polarized with the real part of the electric field as a spatially distributed function along the *x*-axis, the real part of the *y*-polarized output wave will be proportional to the second-order derivative of the input function. Figure 3b depicts the structure of the designed meta-optic system. The system has three layers, each of which has 25 spatially variant unit cells along the *x* direction and is periodic along the *y* direction, with a spacer of 200 nm between adjacent layers. We define Layer 1 on the input side, and Layer 3 for output. To evaluate the performance of the differentiator, we tested it with three different inputs. The three images on the left column of Fig. 2c present the input functions, while the middle and the right columns represent the corresponding target outputs and the simulated outputs of the metasurface, respectively. While there are some discrepancies between the targets and the simulated results of the device, the envelopes share great similarities, which validates the performance of the differentiator. The notable ripples in the simulated output are caused by the limited resolution of the differentiator. Since the meta-optic system consists of only 25 unit cells along the *x* direction for the representation and operation of a continuous function, the computation resolution of the differentiation is limited due to the discretization in the spatial domain. Increasing the amount of unit cells of the metasurface is expected to substantially improve the resultant performance. Apart from the one-dimensional input waves, for two-dimensional images as the input, the differentiator will detect the second-order edges of the image along the *x* direction. If the meta-system based second-order differentiator is designed to be spatially variant in two dimensions, it will be able to distinguish all the second-order edges of an input image. As a result, the devices designed by the AI framework are potentially applicable to all-optical image processing, computing, neural networks, and more.

In the last example of multilayer multifunctional meta-optic system, we will present, to the best of our knowledge, the first space-polarization-wavelength multiplexed metasurface hologram.

This meta-hologram functions at the wavelengths (in the polymer) of both 562 and 659 nm, in both *x* and *y* polarizations, with 9 operational positions along the propagation, and projects a total of 36 holographic images: numerical digits 0-9 and capital letters A-Z. Metasurface has been exploited as one of the most promising media to achieve holography, with high imaging quality and ultrathin thickness[30]. Most of such optical meta-holograms are static and monofunctional, which leads to limited information-storing capacity of the holography. Multiplexed metasurfaces may open up the vistas to encode an enormous amount of information into a single hologram. Some of the recent works have shown the possibility to realize a multiplexed hologram in the spatial, polarization, or wavelength channel, and some have attempted two of the above channels[10,31]. Here, we present an unprecedented hologram that utilizes all three channels, with the largest total amount of displayed images from a single meta-hologram reported to date.

The hologram is neither phase-only nor amplitude-only, but with mixed phase and amplitude information carried in one integrated system. It consists of three layers, with an overall size of 2,000 by 2,000 supercells. The design flow chart is presented in Fig. 4a. We first virtually align all the objective images on axis and parallel to the hologram plane at different distances (100 – 740 μm from the meta-hologram, with 80 μm interval). While the projected function only pertains to the magnitude of light at the generated images, we still need to introduce random phase to the images in order to suppress the crosstalk between different image planes. Next, we apply a Sommerfeld-Fresnel transformation (Huygens propagation) to obtain the amplitude and phase distribution on the hologram plane as derived from the image planes. The final amplitude and phase distribution is the weighted sum of the projections, where the weight is proportional to the square root of the distance from the meta-hologram to a particular image. The weight is introduced to ensure that each image can be recovered by the hologram with both high fidelity and sufficient brightness. To reduce the computational complexity, the objective field amplitude after the hologram is binarized into the values of 0.75 or 0 with a threshold of 0.3. The phase is also discretized into three values within the $2\pi$ period: $-\pi$, $-\frac{1}{3}\pi$ and $\frac{1}{3}\pi$. Therefore, for each polarization at a specific wavelength, there will be four possible amplitude and phase combinations: 0, $-0.75$, $0.75\angle-\frac{1}{3}\pi$, and $0.75\angle\frac{1}{3}\pi$. Since the hologram is designed to work at two different wavelengths (562 nm and 659 nm) and two polarization states (*x* and *y* polarizations), there are totally $(4^2)^2 = 256$ amplitude and phase combinations, which means the same number of unique supercells should be used to constitute the meta-hologram. The detailed structures of the supercells are presented in the Supplementary Information, and Fig. 4b shows a few representative ones. The numbers below each supercell indicate its target responses under *x*- and *y*-polarized incidence at 562 nm ($\lambda_1$) and 659 nm ($\lambda_2$), denoted as $E_{\lambda_1 x}, E_{\lambda_1 y}, E_{\lambda_2 x}, E_{\lambda_2 y}$. Once the 256 supercells are designed by the algorithm, any hologram that operates for both *x* and *y* polarizations at the two wavelengths can be designed in a matter of seconds, since different holograms are simply different maps of the same collection of supercells. In this example, the hologram consists of 2000×2000 supercells, and Fig.4c displays a small portion of the overall structure. We assume the incidence light illuminates from Layer 1, and the output light beyond Layer 3 automatically forms the projected images at the expected locations. Figure 4d shows the 36 simulated holographic images, and the color tones of the images are applied to distinguish the shorter (562 nm, blue) and longer (659 nm, red) operating wavelengths. Most of the holographic images are formed with clear outlines and fine features. The image quality does not degrade at large distances, and the fidelity of images is satisfactory at each polarization and each wavelength.

As shown in Fig. 4d, some of the projected images are not perfect in terms of the sharpness and legibility, which mainly result from crosstalk with other images. There are two major causes of the crosstalk. First, as the projection distance increases, the light propagation will exhibit more of the nature of Fraunhofer diffraction than Fresnel diffraction. As a result, the projected image will mostly represent the Fourier transformation of the input hologram regardless of the distance, rather than the convolution between the hologram and the distance-related transfer function. This is consistent with the observation from Fig. 4d that the blurring or crosstalk of the images becomes more noticeable at longer distances. This dilemma can be settled by adding a Fresnel plate before the hologram to convert all the transformation into the near-field regime[32]. Second, if a projected image is located between two similar images (e.g. "7" between "6" and "8", "P" between "O" and "Q"), it will suffer more from crosstalk than regular cases. Considering the crosstalk stemming from the nonorthogonality between the projected matrices of the images, the image (e.g., "P") sandwiched between two very similar images (e.g., "O" and "Q") will be hardly orthogonal to both, and the crosstalk subsequently arises. To mitigate this shortcoming, the phase randomization process in Fig. 4a can be replaced by a Gram-Schmidt process to guarantee the orthogonality of the projected matrices of the phase-added images. As for the scalability, the 256 unique supercells can be readily adapted to design any space-polarization-wavelength multiplexed holograms for the polarization states and operating wavelengths specified before. Additional examples are presented in the Supplementary Information. It is worth noting that our method can be extended to the design of holograms at more operating wavelengths and image positions. Adding one wavelength with both polarizations will increase the number of unique supercells by 16 times.

In summary, we have proposed a hybrid AI framework for designing highly complicated, multifunctional meta-optic systems. The framework is of high efficiency and accuracy, requires only a small amount of data and a low computational expense, and the algorithm proceeds automatically with no need for human intervention. It is very versatile and can be applied to other configurations through transfer learning. We have presented three examples designed by the AI framework: a polarization-multiplexed dual-functional beam generator, a second order differentiator for all-optical computation, and a space-polarization-wavelength multiplexed hologram. These devices are virtually insoluble through conventional design method, with functions hardly achievable by a single-layer metasurface. The AI framework developed here is also applicable to the design of other photonic components and systems, including photonic crystals, chip-scale silicon devices, and quantum-optical devices. The methodology of our AI-design algorithm is also significant to other disciplines of natural sciences, such as the design of nano materials, searching for new topological insulators, planning of chemical syntheses, prediction of protein structures, and many more. Consolidating the power of AI into scientific research foresees the emergence of novel devices beyond human design capacities, discovers underlying physics from nebulous or counter-intuitive observations, and pushes forward the limits of knowledge as we know today.

**References**

1. Yu, N. & Capasso, F. Flat optics with designer metasurfaces. *Nature Materials* **13**, 139-150 (2014).
2. Bomzon, Z. e., Biener, G., Kleiner, V. & Hasman, E. Space-variant Pancharatnam–Berry phase optical elements with computer-generated subwavelength gratings. *Optics Letters* **27**, 1141-1143 (2002).
3. Aieta, F., Kats, M. A., Genevet, P. & Capasso, F. Multiwavelength achromatic metasurfaces by dispersive phase compensation. *Science* **347**, 1342-1345 (2015).
4. Wang, S. *et al.* A broadband achromatic metalens in the visible. *Nature Nanotechnology* **13**, 227-232 (2018).
5. Jang, M. *et al.* Wavefront shaping with disorder-engineered metasurfaces. *Nature Photonics* **12**, 84-90 (2018).
6. Li, L. *et al.* Machine-learning reprogrammable metasurface imager. *Nature Communications* **10**, 1082 (2019).
7. Brongersma, M. L. & Shalaev, V. M. The case for plasmonics. *Science* **328**, 440-441 (2010).
8. Lin, D., Fan, P., Hasman, E. & Brongersma, M. L. Dielectric gradient metasurface optical elements. *Science* **345**, 298-302 (2014).
9. Wang, B. *et al.* Visible-frequency dielectric metasurfaces for multiwavelength achromatic and highly dispersive holograms. *Nano Letters* **16**, 5235-5240 (2016).
10. Wen, D. *et al.* Helicity multiplexed broadband metasurface holograms. *Nature Communications* **6**, 8241 (2015).
11. Raeker, B. O. & Grbic, A. Compound metaoptics for amplitude and phase control of wave fronts. *Physical Review Letters* **122**, 113901 (2019).
12. Zhou, Y., Zheng, H., Kravchenko, I. I. & Valentine, J. Flat optics for image differentiation. *Nature Photonics* **14**, 316-323 (2020).
13. LeCun, Y., Bengio, Y. & Hinton, G. Deep learning. *Nature* **521**, 436 (2015).
14. Donahue, J. *et al.* Long-term recurrent convolutional networks for visual recognition and description. In *Proceedings of the IEEE Conference on Computer Vision and Pattern Recognition.* 2625-2634 (2015).
15. He, K., Zhang, X., Ren, S. & Sun, J. Deep residual learning for image recognition. In *Proceedings of the IEEE Conference on Computer Vision and Pattern Recognition.* 770-778 (2016).
16. Carleo, G. & Troyer, M. Solving the quantum many-body problem with artificial neural networks. *Science* **355**, 602-606 (2017).
17. Shen, D., Wu, G. & Suk, H.-I. Deep learning in medical image analysis. *Annual Review of Biomedical Engineering* **19**, 221-248 (2017).
18. Peurifoy, J. *et al.* Nanophotonic particle simulation and inverse design using artificial neural networks. *Science Advances* **4**, eaar4206 (2018).
19. Malkiel, I. *et al.* Plasmonic nanostructure design and characterization via Deep Learning. *Light: Science & Applications* **7**, 60 (2018).
20. Liu, D., Tan, Y., Khoram, E. & Yu, Z. Training deep neural networks for the inverse design of nanophotonic structures. *ACS Photonics* **5**, 1365-1369 (2018).
21. Jiang, J. & Fan, J. A. Global optimization of dielectric metasurfaces using a physics-driven neural network. *Nano Letters* **19**, 5366-5372 (2019).
22. Ma, W., Cheng, F., Xu, Y., Wen, Q. & Liu, Y. Probabilistic representation and inverse design of metamaterials based on a deep generative model with semi-supervised learning strategy. *Advanced Materials* **31**, 1901111 (2019).

23. Li, Y. *et al.* Self-Learning Perfect Optical Chirality via a Deep Neural Network. *Physical Review Letters* **123**, 213902 (2019).
24. Molesky, S. *et al.* Inverse design in nanophotonics. *Nature Photonics* **12**, 659 (2018).
25. Liu, Z. *et al.* Compounding meta-atoms into metamolecules with hybrid artificial intelligence techniques. *Advanced Materials* **32**, 1904790 (2020).
26. Goodfellow, I. *et al.* Generative adversarial nets. In *Advances in Neural Information Processing Systems.* 2672-2680 (2014).
27. Stanley, K. O. Compositional pattern producing networks: A novel abstraction of development. *Genetic Programming and Evolvable Machines* **8**, 131-162 (2007).
28. Silva, A. *et al.* Performing mathematical operations with metamaterials. *Science* **343**, 160-163 (2014).
29. Zhu, T. *et al.* Plasmonic computing of spatial differentiation. *Nature Communications* **8**, 15391 (2017).
30. Ni, X., Kildishev, A. V. & Shalaev, V. M. Metasurface holograms for visible light. *Nature Communications* **4**, 2807 (2013).
31. Huang, L. *et al.* Broadband hybrid holographic multiplexing with geometric metasurfaces. *Advanced Materials* **27**, 6444-6449 (2015).
32. Makey, G. *et al.* Breaking crosstalk limits to dynamic holography using orthogonality of high-dimensional random vectors. *Nature Photonics* **13**, 251-256 (2019).

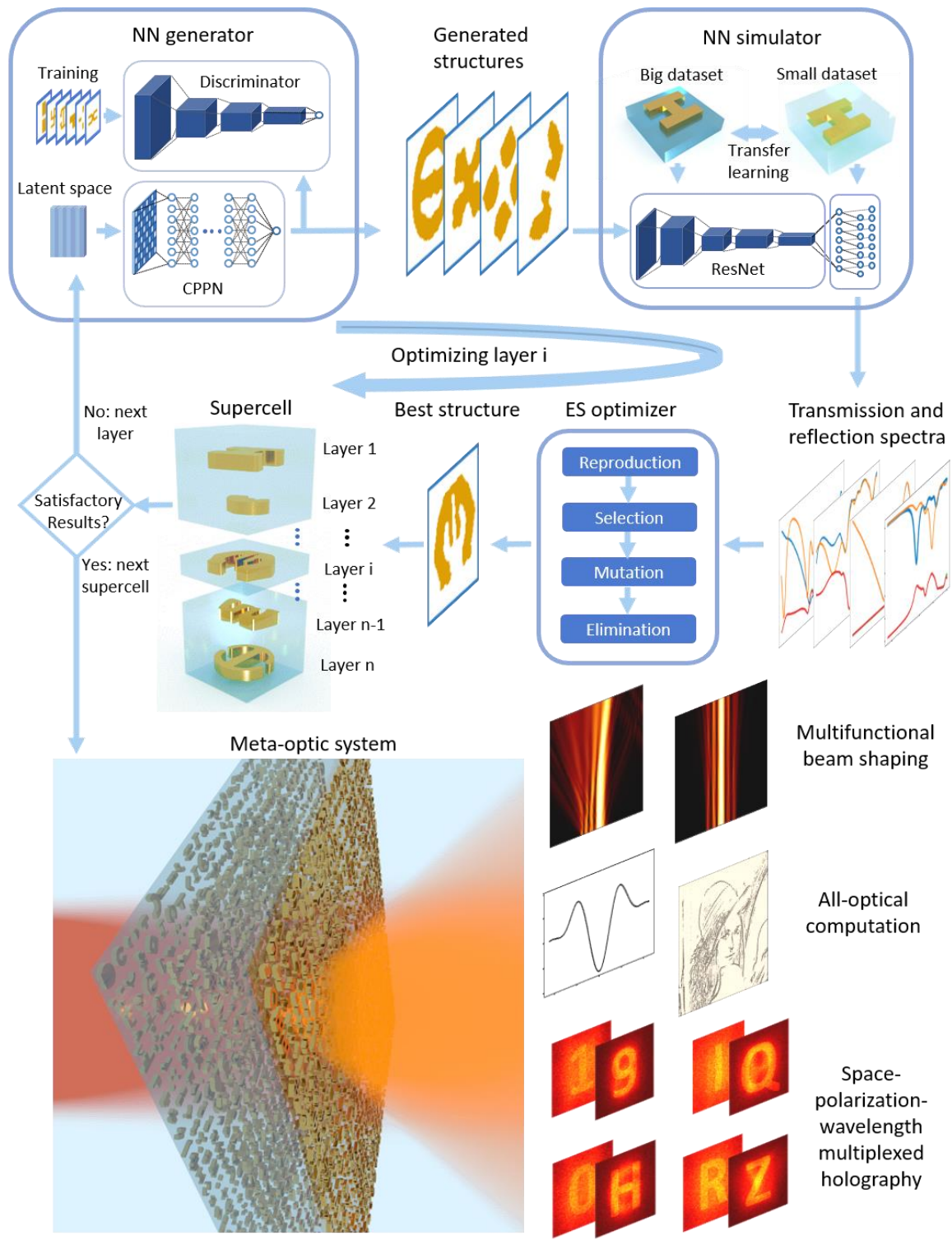

**Figure 1.** Description of the modules of the artificial intelligence (AI) framework. The design flow goes through the neural network (NN) generator, the NN simulator, and the evolutionary strategy (ES) optimizer in a loop. The generator is a Compositional-Pattern Generating Network (CPPN) of a Generative Adversarial Network (GAN). The simulator is trained sequentially through transfer learning. The workflow proceeds as following: the generator first

decodes a batch of latent vectors and produces corresponding candidate nanostructures, followed by the NN simulator computing the transmittance and reflectance spectra of the nanostructures. Next, the ES optimizer selects the structures with good performance, refines and evolves their latent vectors, and sends the latent vectors back to the generator to get a better batch of nanostructures. Different layers at the same position can be collectively viewed as a supercell. One layer of the supercell is optimized in every single iteration, and the design proceeds in a round-robin fashion until the overall performance of the supercell is satisfactory. The design of the meta-optic system terminates when all the spatially variant supercells are obtained. The meta-optic system can be designed as a multifunctional optical device, for applications such as multifunctional beam shaping, and space-wavelength-polarization multiplexed holography. The meta-optic system is also applicable to all-optical computation.

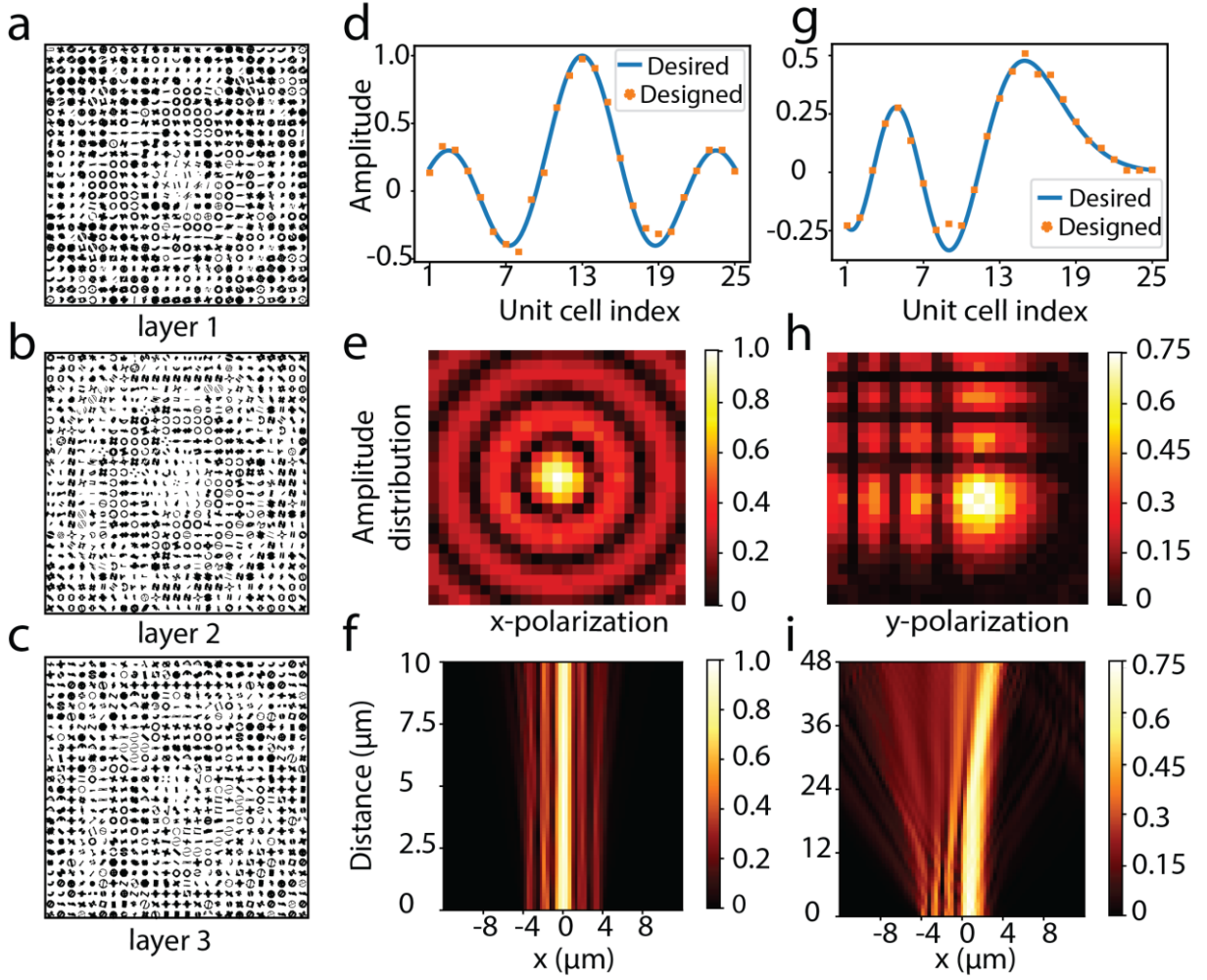

**Figure 2.** The AI-designed meta-optic system as a polarization-multiplexed dual-functional beam generator. When the incident light is *x*-polarized, the transmitted *x*-polarized wave forms a Bessel beam, while *y*-in-*y*-out response is an Airy beam. (a-c) Detailed structure of the meta-optic system. The system has three layers (a-c), each layer consists of 25×25 unit cells. The light incidence is on Layer 1 side and output on Layer 3 side. (d and g) Desired amplitude distributions (blue) and simulated amplitudes (orange) of the meta-optic system on the central axis, for the *x*-polarized and the *y*-polarized incidence, respectively. (e and h) Simulated amplitude distributions of the light field on the output plane, for the *x*- and *y*-polarized incidence, respectively. (f and i) Simulated amplitude distributions of the transmitted light along the propagation *z* direction, for the *x*- and *y*-polarized incidence, respectively.

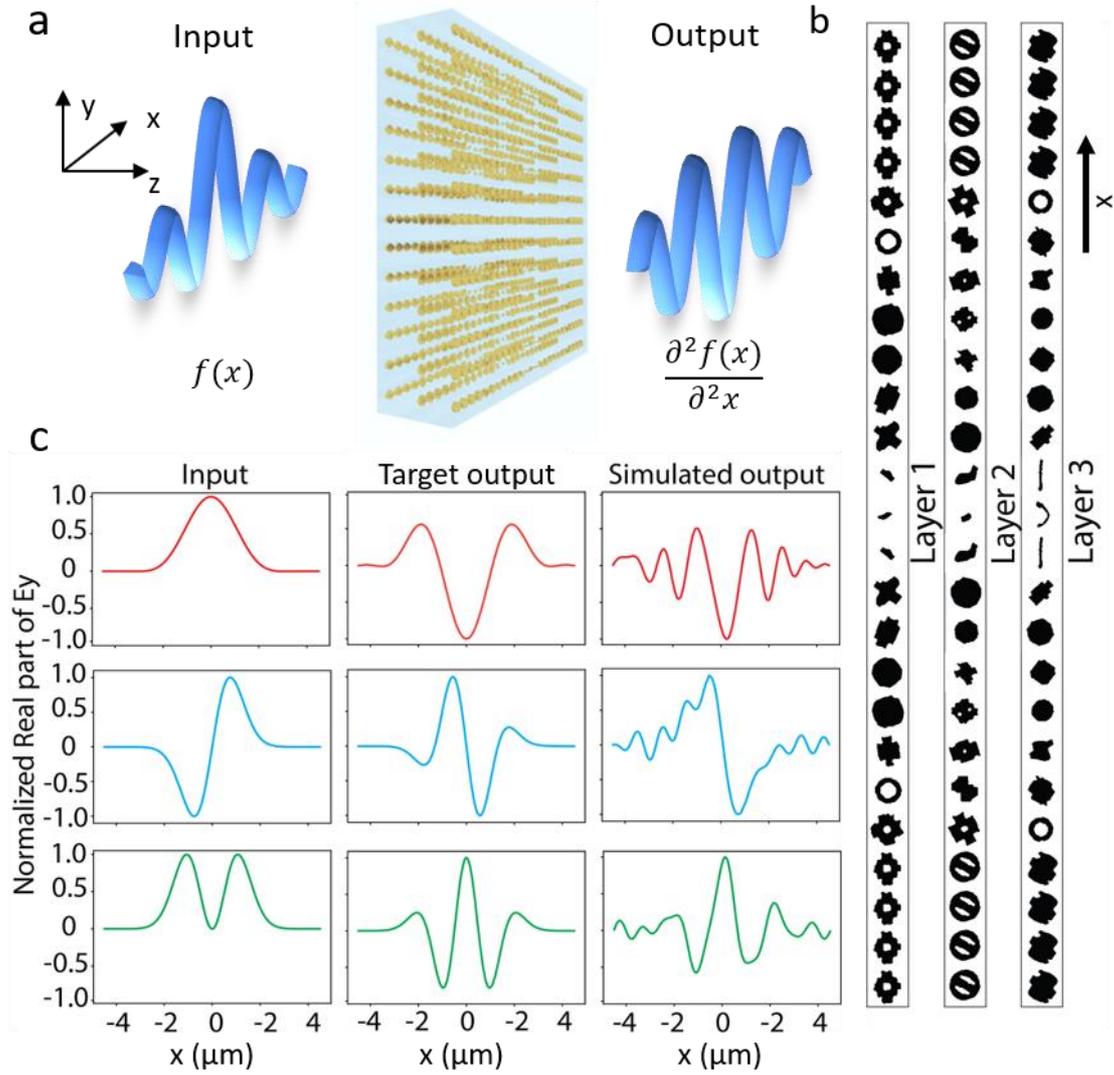

**Figure 3.** The AI-designed meta-optic system for second-order differentiation. (a) Schematic of the function of the second-order differentiator. When the input is a wave with spatially variant real part along the *x* direction, the real part of the output wave will be the second-order derivative of the input. (b) Detailed structure of the meta-optic system. Each layer of the tri-layered system has 25 spatially variant unit cells along the *x* direction, and is periodic along the *y* direction. The incidence is on Layer 1 side and the output is on Layer 3 side. (c) Three examples to test the functionality of the differentiator. The left column is three examples of the input, while the middle and right columns are the target output and the actual output of the meta-optic second-order differentiator, respectively. Since the differentiator has only 25 unit cells along the *x* direction, the resolution of the differentiation is limited, which leads to the ripples in the simulated output.

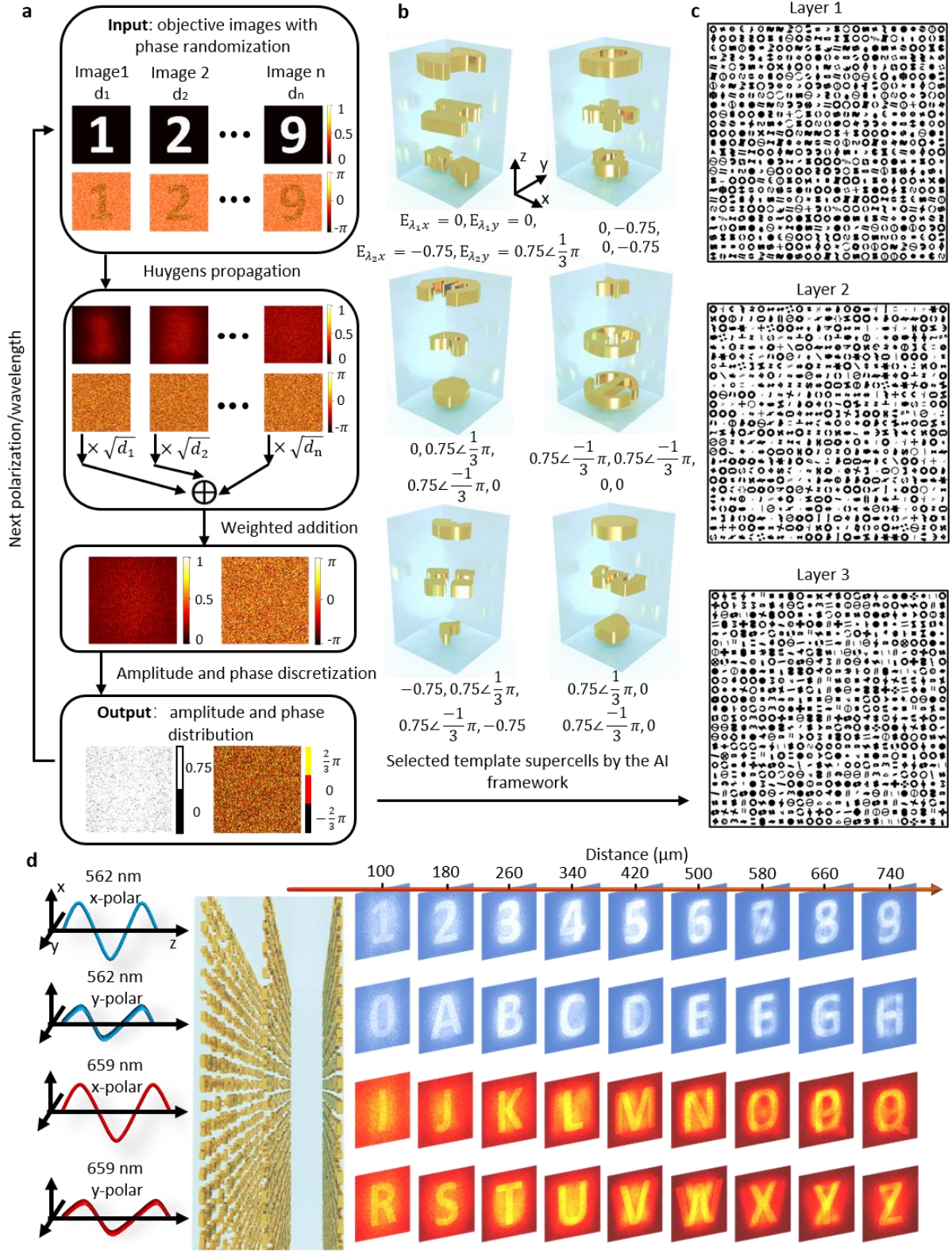

**Figure 4.** AI-designed space-polarization-wavelength multiplexed hologram. The hologram can project 36 different holographic images in total. (a) The flowchart to obtain the amplitude-and-phase distributions of the design objective. The first step is to pin down the target images projected by the hologram and their distances to the hologram plane, then random phases are added to each of them. Next, their projected amplitude-and-phase distributions on the hologram plane are computed by the Huygens propagation, and the weighted superposition is calculated, then the amplitude is normalized to between 0 and 1. Lastly, the amplitude and the phase are discretized. The hologram functions at two wavelengths and two polarizations, which means the process should be carried out four times to get all the distributions. (b) Some examples of the designed supercells. Their target responses at *x* and *y*-polarized incidence at 562 nm ($\lambda_1$) and 659 nm ($\lambda_2$) in the polymer are denoted as $E_{\lambda_1 x}, E_{\lambda_1 y}, E_{\lambda_2 x}, E_{\lambda_2 y}$. (c) A part of the designed hologram. The total hologram consists of 2,000 by 2,000 tri-layered supercells, and here we display a locality of 25 by 25 supercells. The incidence is on Layer 1 side and output on Layer 3 side. (d) Simulated performances of the hologram. The hologram works at both *x* and *y* polarizations, 659 and 562 nm, and the projected images are at 9 distances to the hologram, from 100 µm to 740 µm. The projected 36 images are the numerical digits 0 to 9, and capital letters A to Z.